\documentclass[12pt]{article}
\usepackage{amsfonts}
\usepackage{amsmath}
\usepackage{amsthm}
\usepackage{graphicx}
\usepackage[compress]{cite}
\usepackage{float}  
\numberwithin{equation}{section}
\title{Dbar dressing method to nonlinear Schr\"{o}dinger equation with nonzero boundary conditions}
\date{}
\author{Junyi Zhu$^{a}$\thanks{jyzhu@zzu.edu.cn}, Xueling Jiang$^{b}$ and Xueru Wang$^{a}$\thanks{510722521@qq.com}\\
\scriptsize{\sl $^{a}$ School of Mathematics and Statistics, Zhengzhou University, Zhengzhou, Henan 450001, China}\\
\scriptsize{\sl $^{b}$ Zhengzhou E-Commerce Vocational College, Zhengzhou, Henan 450048, China}}

\begin{document}
\maketitle {}
\begin{abstract}
The Dbar dressing method is extended to study the focusing/defocusing nonlinear Schr\"odinger (NLS) equation with nonzero boundary condition. A special type of complex function is considered. The function is meromorphic outside an annulus with center 0 and satisfies a local Dbar problem inside the annulus. The theory of such function is extended to construct the Lax pair of the NLS equation with nonzero boundary condition. In this procedure, the relation between the NLS potential and the solution of the Dbar problem is established. A certain distribution for the Dbar problem is introduced to obtain the focusing/defocusing NLS equation and the conservation laws. The explicit solutions of the focusing/defocusing NLS equation with nonzero boundary condition are given from special distributions.

{\bf Keywords}: Dbar dressing method, nonlinear Schr\"odinger equation, nonzero boundary condition
\end{abstract}

\section{Introduction}

Nonlinear integrable equations with nonzero boundary condition (NZBC) have been well studied. Among the methods, the inverse scattering transform or the Riemann-Hilbert problem play an important role \cite{spj37-823,jpsj44-1968,jpsj44-1722,pre69-066604,jmp47-063508,ip23-1711,sam126-245,sam131-1,jmp55-031506,jmp55-101505,sam132-138,non28-3101,jmp56-071505,cm651-157,
jnmp22-233,sjma47-706,pd333-117,cmp348-475,ijam82-131,jmp59-011501,pd368-22,non31-5385,cnsns67-555,jmaa474-452,cnsns80-104927,pd402-132170,zamp71-149,cmp382-87,cmp382-585,sam146-371}.
It is worth noting that Jaulent, Manna, et al. introduced the spatial transform method based on certain Dbar equation to study the integrable systems, such as KdV, Toda and AKNS hierarchy \cite{pla117-62,ip2-L35,ip3-L13,ip4-123}.

The $\bar\partial$ (Dbar) problem is a powerful tools to study the nonlinear integrable equations, such as multidimensional equations \cite{faa19-89,pd18-242,jpa21-L537,ip5-87,KBG1993,ip9-391,sam90-189,jpa27-4619,jpa29-3617,jmp38-6382,jpa32-369,jpa34-1837,aml113-106836},
differential-difference equations \cite{cmp196-1,Santini2003,jmp56-103507,aml113-106836},
(1+1) dimensional equations \cite{ip5-87,D-L2007,jpa46-035204,mpag17-49,jmaa426-783,aml66-47,aml110-106589}.
The Dbar-steepest descent method is developed to study the the asymptotic behavior \cite{imrp2006-48673,imrn2008-rnn075,1912.10358}. The Dbar problem can also be used to consider the well-posedness of integrable equations \cite{anmath143-71,im220-395}.
To our knowledge, very few of the nonlinear integrable equations with NZBCs are considered by the Dbar problem. we note that multi-lump solutions of {KP} equation with integrable boundary $u_y|_{y=0}=0$ via $\bar\partial$-dressing method were given in \cite{pd414-132740}.

In this paper, we give a different view to know about the nonlinear integrable equations with NZBCs. As an example, we extend the Dbar approach to discuss the focusing and defocusing nonlinear Schr\"odinger (NLS) equation with NZBC.  The associated theory is developed, and can also be used to discuss other nonlinear integrable equations with NZBCs. For convenience, we consider the NLS equation with nonzero boundary condition in the following form \cite{F-T1987}
\begin{equation}\label{a1}
iq_{t}+q_{xx}-2\nu(|q|^{2}-q_{0}^{2})q=0, \quad \nu=\pm1,
\end{equation}
and
\begin{equation}\label{a2}
q(x,t)\rightarrow \rho, \quad |x|\to \infty,
\end{equation}
where $\rho$ is a constant and $|\rho|=q_{0}\neq0$.
Equation (\ref{a1}) is the compatibility condition of the linear system
\begin{equation}\label{a3}
 \varphi_x=U\varphi, \quad \varphi_t=V\varphi,
\end{equation}
where
\begin{equation}\label{a4}
 U=\left(\begin{matrix}
 ik&q\\
 \nu\bar{q}&-ik
 \end{matrix}\right), \quad
 V=\left(\begin{matrix}
 -2ik^2-i\nu(|q|^2-q_0^2)&-2kq+iq_x\\
 -2k\nu\bar{q}-i\nu\bar{q}_x&2ik^2+i\nu(|q|^2-q_0^2)
 \end{matrix}\right).
\end{equation}
It is noted that the eigenvalues of the matrix $U_0=U(q=\rho)$ have double branches, and the associated spectral space for the nonlinear Schr\"odinger (NLS) equation with NZBC is multi-sheeted Riemann surface \cite{jmp55-031506,sam131-1}. To use the Dbar approach solving the NLS equation with nonzero boundary condition, one needs to transform the multi-sheeted Riemann surface into a Riemann sphere. This can be done by introducing the the uniformization variable $z$ defined by $z=k+\lambda$ and
\begin{equation}\label{a5}
\lambda(z) =\frac{1}{2}(z-\nu\frac{q_{0}^{2}}{z}), \quad k(z)=\frac{1}{2}(z+\nu\frac{q_{0}^{2}}{z}).
\end{equation}
Hence, the eigenfunction of the spectral problem (\ref{a3}) as $q=\rho$ can be given as
\begin{equation}\label{a6}
 \left(I+\frac{i}{z}\sigma_3Q_0\right)e^{i\theta(z;x,t)\sigma_{3}},
\end{equation}
where
\begin{equation}\label{a7}
\begin{aligned}
Q_0=\left(\begin{matrix}
0&\rho\\
\nu\bar\rho&0
\end{matrix}\right),\quad \theta (z;x,t)=\lambda(z) (x-2k(z)t).
\end{aligned}
\end{equation}

In the following, we consider the Dbar problem in the extended complex $z$ plane. To do this, we construct an annulus with center at 0, that is, 0 and $\infty$ are outside the annulus. In Section 2, we introduce a special complex function which satisfies a Dbar problem in the annulus, and is meromorphic outside the annulus. Thus the Laurent series near the points 0 and $\infty$ play the role of non-canonical normalization conditions to the Dbar problem. The Dbar problem with normalization conditions is equivalent to an inhomogeneous integral equation, and the inhomogeneous terms are given by the normalization conditions. We present the following theorem proved in Appendix to fullfill the Dbar dressing.

{\bf Theorem }~~ {\sl Suppose that $f(z)$ admits $\bar\partial f(z)\neq0$ in $\mathbb{C}^0=\mathbb{C}\setminus\{0\}$. 
If $f(z)$ satisfies the following asymptotic behaviors in $\mathbb{C}^*=\mathbb{C}\cup\{\infty\}$
\begin{equation}\label{a11}
\begin{aligned}
f(z)= \sum\limits_{j=1}^m\frac{a_{m-j}}{z^{j}}+O(1),\quad z\to0,\\
f(z)= \sum\limits_{j=0}^nb_jz^{j}+O(1/z),\quad z\to\infty,
\end{aligned}
\end{equation}
then for the circles $\Gamma_R=\{z:|z|=R>0\}$ and $\Gamma_\varepsilon=\{z:|z|=\varepsilon>0\}$
\begin{equation}\label{a12}
\frac{1}{2\pi i}\oint_{\Gamma_R+\Gamma_\varepsilon^-}\frac{f(k)}{k-z }{\rm d}k
=\sum\limits_{j=0}^nb_jz^j+\sum\limits_{j=1}^m\frac{a_{m-j}}{z^{j}}, \quad \varepsilon<|z|<R,
\end{equation}
where $R\to\infty$ and $\varepsilon\to0$.
}

The Dbar dressing method is based on the hypothesis that the homogeneous integral equation has only zero solution.
To establish the relation between the NLS potential and the solution of the Dbar problem, we construct, in Section 3, the Lax pair of the NLS equation with NZBC. To this end, it is important to find two sets of operator which have same normalization conditions.

A special distribution (or spectral transform matrix) for the Dbar problem is introduced in Section 4 to construct the NLS equation under the nonzero boundary conditions and the conservation laws. The determinant of the associated eigenfunction (or the solution of the Dbar problem) is shown to be analytic in the annulus. In this procedure, we introduce a symmetry matrix function about the eigenfunction, and give its evolution equation in terms of the Lax pair. By substituting the expansion of the symmetry matrix function into the evolution equation and taking the $O(z^l)$ terms, we find the NLS equation with nonzero boundary condition from the off-diagonal parts, and the conservation laws from the diagonal parts. We note that the AKNS hierarchy and infinite conservation laws are shown in \cite{pla117-62}.

In Section 5, the explicit solutions of the focusing and defocusing NLS equation with NZBC are obtained from two special distributions, which make sure the small norm of the operator in the integral equation associated with the Dbar problem. $N$-soliton solutions of the NLS equation with NZBC are given, and for the defocusing NLS equation, dark one-soliton and dark two-soliton are presented. We show that that the collision angle of dark two-soliton in $x$-$t$ plane is determined not only by the eigenvalues but also by the boundary condition.

The conclusions are given in Section 6. At last, the theory of the normalization part of the associated eigenfunctions are presented in the Appendixes.

\setcounter{equation}{0}
\section{Dbar-problem with non-canonical normalization conditions}
We consider the following $\bar{\partial}$(Dbar)-problem
\begin{equation}\label{b1}
\bar{\partial}\chi(z;x,t):=\frac{\partial\chi(z;x,t)}{\partial\bar{z}}= \chi(z;x,t)r(z), \quad z\in\mathbb{C}^0,
\end{equation}
where $\chi(z;x,t),r(z)$ are $2\times2$ matrices, the distribution $r(z)$ is independent of $x$ and $t$.
To study the NLS equation with NZBC, we introduce the following normalization condition
\begin{equation}\label{b2}
\begin{aligned}
\chi(z;x,t)&\sim e^{i\theta(z;x,t) \sigma_{3}}, \quad z\rightarrow \infty,\\
\chi(z;x,t)&\sim \frac{i}{z}\sigma_{3}Q_0 e^{i\theta(z;x,t)\sigma_{3}},\quad z\rightarrow 0.
\end{aligned}
\end{equation}
We note that in the procedure of the inverse scattering transform, one can introduce the Jost functions which tend to (\ref{a6}) as $|x|\to\infty$. It is remarked that the condition (\ref{a6}) implies the normalization condition (\ref{b2}) for the Dbar problem.

For simplicity, we introduce a new function
\begin{equation}\label{b4}
\hat{\chi}(z;x,t)=\chi(z;x,t)e^{-i\theta(z;x,t)\sigma_{3}},
\end{equation}
then it satisfies the asymptotic behavior
\begin{equation}\label{b5}
\hat{\chi}(z;x,t)\sim I, \quad z\rightarrow\infty,
\end{equation}
\begin{equation}\label{b6}
\hat{\chi}(z;x,t)\sim \frac{i}{z}\sigma_{3}Q_0, \quad z\rightarrow0.
\end{equation}
and the generalised Cauchy integral formula
\begin{equation}\label{b7}
\hat{\chi}(z)=\frac{1}{2\pi i}\int_{\Gamma_R+\Gamma_{\varepsilon}^{-}}\frac{\hat{\chi}(k)}{k-z}\mathrm{d}k+\frac{1}{2\pi i}\iint_{\varepsilon<|k|<R}\frac{\bar\partial\hat{\chi}(k)}{k-z}\mathrm{d}k\wedge \mathrm{d}{\bar{k}},
\end{equation}
where $\Gamma_R$ and $\Gamma_\varepsilon$ are oriented circle with center at origin of $z$ plane and radius $R$ and $\varepsilon$, respectively. Here, $R\to\infty$ and $\varepsilon\to0$. For simplicity, we define the first Cauchy integral on the right hand side of (\ref{b7}) as the normalization part of $\chi$, and denote it as $\mathcal{N}(\chi)$, that is
\begin{equation}\label{b8a}
 \mathcal{N}(\chi)=\frac{1}{2\pi i}\int_{\Gamma_R+\Gamma_{\varepsilon}^{-}}\frac{\hat{\chi}(k)}{k-z}\mathrm{d}k.
\end{equation}
Then for $\chi$ in (\ref{b2}), we find from the {\bf Theorem} that
\begin{equation}\label{b12}
 \mathcal{N}(\chi)=I+\frac{i}{z}\sigma_{3}Q_0.
\end{equation}

As $R\rightarrow \infty$ and $\varepsilon\rightarrow 0$, (\ref{b7}) reduces to
\begin{equation}\label{b8}
\hat{\chi}(z)=I+\frac{i}{z}\sigma_{3}Q_0+J\hat{\chi}(z),
\end{equation}
where
\begin{equation}\label{b9}
J\hat{\chi}(z)=\frac{1}{2\pi i}\iint_{\mathbb{C}^0}\frac{\hat{\chi}(k)e^{i\theta(k)\sigma_{3}}r(k)e^{-i\theta(k)\sigma_{3}}}{k-z}\mathrm{d}{k}\wedge \mathrm{d}{\bar{k}}.
\end{equation}

It is important to assume that the homogeneous equation of (\ref{b8}) only has zero solution, that is,
\begin{equation}\label{b10}
 (I-J)f=0 \Rightarrow f=0.
\end{equation}
It is valid for small norm of the operator $J$.

Now we introduce the following solution space of the Dbar-problem (\ref{b1}) as
\begin{equation}\label{b11}
\mathcal{F}=\{\chi(z;x,t)|\bar{\partial}\chi(z;x,t)= \chi(z;x,t)r(z), \quad z\in\mathbb{C}^0\}.
\end{equation}

In particular, let $\psi(x,;z)\in\mathcal{F}$ and $\mathcal{N}(\psi)=I+\frac{i}{z}\sigma_3Q_0$. Note that the distribution $r(z)$ is independent of the variables $x$ and $t$.
To study the NLS equation with NZBC via the Dbar-problem (\ref{b1}), one needs to introduce certain constraint in the physic space. Here, we suppose
\begin{equation}\label{b13}
 \psi(z;x,t)\sim\left(I+\frac{i}{z}\sigma_3Q_0\right)e^{i\theta(z;x,t)\sigma_{3}}, \quad |x|\to\infty,
\end{equation}
and
\begin{equation}\label{b13a}
 \psi(z;x,t)=\frac{i}{z}\psi(\nu\frac{q_{0}^{2}}{z};x,t)\sigma_{3}Q_{0}.
\end{equation}
Then from (\ref{b8}), we know that $\psi(z;x,t)$ has the following asymptotic behaviors
\begin{equation}\label{b14}
\psi(z;x,t)=\left(I+\sum_{l=1}^\infty a_l(x,t)z^{-l}\right)e^{i\theta(z;x,t)\sigma_{3}}, \quad z\to\infty,
\end{equation}
\begin{equation}\label{b15}
 \psi(z;x,t)=\left(\sum_{m=-1}^\infty b_m(x,t)z^{m}\right)e^{i\theta(z;x,t)\sigma_{3}}, \quad z\to0,
\end{equation}
where
\begin{equation}\label{b16}
a_l(x,t)=\delta_{l,1}\cdot i\sigma_{3}Q_{0}-\frac{1}{2\pi i}\iint_{D}\psi(z;x,t)r(z)e^{-i\theta(z)\sigma_{3}}z^{l-1}\mathrm{d}z\wedge \mathrm{d}{\bar{z}},
\end{equation}
and
\begin{equation}\label{b17}
b_m(x,t)=\left\{\begin{array}{ll}\delta_{m,0}+\frac{1}{2\pi i}\iint_{D}\psi(z;x,t)r(z)e^{-i\theta(z)\sigma_{3}}z^{-m-1}\mathrm{d}z\wedge \mathrm{d}{\bar{z}},&m\geq0,\\
i\sigma_{3}Q_{0},&m=-1.
\end{array}\right.
\end{equation}
It is remarked that the coefficients $a_l(x,t)$ and $b_m(x,t)$ are not independent in terms of the symmetry condition (\ref{b13a}). In fact, 
\begin{equation}\label{b20}
  b_{-1}=i\sigma_3Q_0, \quad b_{m-1}(x,t)=\frac{i}{\nu^mq_0^{2m}}a_m(x,t)\sigma_3Q_0.
\end{equation}

\section{Dbar dressing for NLS equation with NZBC}
We note that equations (\ref{b10}) and (\ref{b12}) imply that, for $\chi_1,\chi_2\in\mathcal{F}$,
\begin{equation}\label{c1}
 \mathcal{N}(\chi_1)=\mathcal{N}(\chi_2) \Leftrightarrow\chi_1=\chi_2.
\end{equation}
This result can be used to construct the Lax pair of NLS equation (\ref{a1}). In fact, Since the distribution $r(z)$ is independent of the variables $x$ and $t$, we know that $\alpha(z;x,t)\partial_x\psi+\beta(z;x,t)\partial_t\psi+A(z;x,t)\psi\in\mathcal{F}$, if $\psi\in\mathcal{F}$. Here $\alpha(z;x,t),\beta(z;x,t)$ and $A(z;x,t)$ are some $2\times2$ matrices.

To obtain the spatial linear spectral problem, a little manipulation is needed. Let $\psi\in\mathcal{F}$. Here and after, $\psi=\psi(z;x,t)$. From (\ref{b14}), we find, at $z\to\infty$, that
\begin{equation}\label{c2}
\begin{aligned}
\psi_x=&\left(\frac{i}{2}\sigma_{3}z+\frac{i}{2}a_{1}\sigma_{3}+\frac{i}{2}\sum_{l=1}^{\infty}a_{l+1}\sigma_{3}z^{-l}\right.\\
&\quad\left.+\frac{iq_{0}^{2}}{2}(\frac{\sigma_{3}}{z}+\sum_{l=2}^{\infty}a_{l-1}\sigma_{3}z^{-l})+\sum_{l=1}^{\infty}a_{l,x}z^{-l}\right)e^{i\theta(z)\sigma_{3}},
\end{aligned}
\end{equation}
\begin{equation}\label{c3}
ik\sigma_{3}\psi=\frac{i}{2}\left(\sigma_{3}z+\sigma_{3}a_{1}+\sum_{l=1}^{\infty}\sigma_{3}a_{l+1}z^{-l}-q_{0}^{2}(\frac{\sigma_{3}}{z}+\sum_{l=2}^{\infty}\sigma_{3}a_{l-1}z^{-l})\right)e^{i\theta(z)\sigma_{3}},
\end{equation}
\begin{equation}\label{c4}
-\frac{i}{2}[\sigma_{3},a_{1}]\psi=\frac{i}{2}\left(-[\sigma_{3},a_{1}]-\sum_{l=1}^{\infty}[\sigma_{3},a_{1}]a_{l}z^{-l}\right)e^{i\theta(z)\sigma_{3}},
\end{equation}
where $k=k(z)$ is defined in (\ref{a5}), then
\begin{equation}\label{c5}
\psi_x=\left(\frac{i}{2}\sigma_{3}z+\frac{i}{2}a_{1}\sigma_{3}+O(\frac{1}{z})\right)e^{i\theta(z)\sigma_{3}}, \quad z\rightarrow\infty,
\end{equation}
\begin{equation}\label{c6}
\begin{aligned}
ik\sigma_{3}\psi-\frac{i}{2}[\sigma_{3},a_{1}]\psi
&=\left(\frac{i}{2}\sigma_{3}z+\frac{i}{2}a_{1}\sigma_{3}+O(\frac{1}{z})\right)e^{i\theta(z)\sigma_{3}},\quad z\rightarrow\infty,
\end{aligned}
\end{equation}
with $[\sigma_{3},a_{1}]=\sigma_3a_1-a_1\sigma_3$. Equations (\ref{c5}) and (\ref{c6}) imply that the Laurent series of $\partial_{x}\psi$ and $ik\sigma_{3}\psi-\frac{i}{2}[\sigma_{3},a_{1}]\psi$ at $z\to\infty$ share the same principal part.

Similarly, from (\ref{b15}), we know that, as $z\to0$
\begin{equation}\label{c7}
\begin{aligned}
\psi_x=\left({b_{-1,x}z^{-1}-\frac{i}{2}\nu q_{0}^{2}(b_{0}\sigma_{3}z^{-1}+b_{-1}\sigma_{3}z^{-2})+\sum_{m=0}^{\infty}b_{m,x}z^{m}}\right.\\
\left.{+\frac{i}{2}\sum_{m=0}^{\infty}b_{m-1}\sigma_{3}z^{m}-\frac{i}{2}\nu q_{0}^{2}\sum_{m=0}^{\infty}b_{m+1}\sigma_{3}z^{m}}\right)e^{i\theta(z)\sigma_{3}},
\end{aligned}
\end{equation}
\begin{equation}\label{c8}
\begin{aligned}
ik\sigma_{3}\psi=&\left(\frac{i}{2}\nu q_{0}^{2}(\sigma_{3}b_{-1}z^{-2}+\sigma_{3}b_{0}z^{-1})+\frac{i}{2}\sum_{m=0}^{\infty}\sigma_{3}b_{m-1}z^{m}\right.\\
&\quad\left.+\frac{i}{2}\nu q_{0}^{2}\sum_{m=0}^{\infty}\sigma_{3}b_{m+1}z^{m}\right)e^{i\theta(z)\sigma_{3}},
\end{aligned}
\end{equation}
\begin{equation}\label{c9}
\begin{aligned}
&-\frac{i}{2}\nu q_{0}^{2}(\sigma_{3}b_{0}+b_{0}\sigma_{3})b_{-1}^{-1}\psi=\left(-\frac{i}{2}\nu q_{0}^{2}(\sigma_{3}b_{0}+b_{0}\sigma_{3})z^{-1}\right.\\
&\qquad\left.-\frac{i}{2}\nu q_{0}^{2}(\sigma_{3}b_{0}+b_{0}\sigma_{3})b_{-1}^{-1}\sum_{m=0}^{\infty}b_{m}z^{m}\right)e^{i\theta(z)\sigma_{3}},
\end{aligned}
\end{equation}
and further
\begin{equation}\label{c10}
\psi_x=\left(-\frac{i}{2}\nu q_{0}^{2}(b_{-1}\sigma_{3}z^{-2}+b_{0}\sigma_{3}z^{-1})+O(1)\right)e^{i\theta(z)\sigma_{3}}, \quad z\to0,
\end{equation}
\begin{equation}\label{c11}
\begin{aligned}
&ik\sigma_{3}\psi-\frac{i}{2}\nu q_{0}^{2}(\sigma_{3}b_{0}+b_{0}\sigma_{3})b_{-1}^{-1}\psi\\
&\quad=\left(-\frac{i}{2}\nu q_{0}^{2}(b_{-1}\sigma_{3}z^{-2}+b_{0}\sigma_{3}z^{-1})+O(1)\right)e^{i\theta(z)\sigma_{3}}, \quad z\to0,
\end{aligned}
\end{equation}
which implies that the Laurent series of $\partial_{x}\psi$ and $ik\sigma_{3}\psi+\frac{i}{2}q_{0}^{2}(\sigma_{3}b_{0}+b_{0}\sigma_{3})b_{-1}^{-1}\psi$ at $z=0$ share the same principal part.

Now, using the relation (\ref{b20}), we find that the coefficient of second item on the left hand side of (\ref{c11})
is equivalent to that of (\ref{c6}),
\begin{equation}\label{c12}
\frac{i}{2}\nu q_{0}^{2}(\sigma_{3}b_{0}+b_{0}\sigma_{3})b_{-1}^{-1}
=\frac{i}{2}[\sigma_{3},a_{1}].
\end{equation}

Since $\partial_x\psi$ defined by (\ref{c5}) and (\ref{c10}) belongs to the space $\mathcal{F}$, then, from (\ref{b8a}) and the {\bf Theorem}, we find
\begin{equation}\label{c13}
\mathcal{N}(\psi_x)=\frac{i}{2}\sigma_{3}z+\frac{i}{2}a_{1}\sigma_{3}+\frac{i}{2}q_{0}^{2}(b_{0}\sigma_{3}z^{-1}+b_{-1}\sigma_{3}z^{-2}).
\end{equation}
Similarly, for $ik\sigma_{3}\psi-\frac{i}{2}[\sigma_{3},a_{1}]\psi\in\mathcal{F}$ in terms of (\ref{c6}) and (\ref{c11}), we get
\begin{equation}\label{c14}
 \mathcal{N}(\psi_x)=\mathcal{N}(ik\sigma_{3}\psi-\frac{i}{2}[\sigma_{3},a_{1}]\psi),
\end{equation}
which gives the spatial linear spectral problem
\begin{equation}\label{c15}
 \psi_x=ik\sigma_3\psi+Q\psi, \quad Q=-\frac{i}{2}[\sigma_{3},a_{1}],
\end{equation}
in view of the identity (\ref{c1}).

Substituting the expansion (\ref{b14}) into (\ref{c15}) and taking the $O(z^{-1})$ terms, we get the following equation
\begin{equation}\label{c16}
 -\frac{i}{2}a_{2}\sigma_{3}=-i\nu q_{0}^{2}\sigma_{3}-\frac{i}{2}\sigma_{3}a_{2}+a_{1,x}-Qa_{1},
\end{equation}
which can also be derived from equations (\ref{c2})-(\ref{c4}).
Equations (\ref{c15}) and (\ref{c16}) imply that
\begin{equation}\label{c17}
a_{1,x}=i\sigma_{3}(Q_{x}-Q^{2}+\nu q_{0}^{2}).
\end{equation}

Similarly, substituting the expansion (\ref{b15}) into (\ref{c15}), and taking the $O(z^0)$ terms, we get
\begin{equation}\label{c18}
\frac{i}{2}\nu q_{0}^{2}b_{1}\sigma_{3}=b_{0,x}+Q_{0}-\frac{i}{2}\nu q_{0}^{2}\sigma_{3}b_{1}-Qb_{0}.
\end{equation}
Here $b_{-1}=i\sigma_3Q_0$ has been used.

Next, we will derive the temporal linear spectral problem. As $z\to\infty$, from (\ref{c14}), we have
\begin{equation}\label{b40}
\begin{aligned}
\psi_t=&\left({-\frac{i}{2}\sigma_{3}z^{2}-\frac{i}{2}a_{1}\sigma_{3}z-\frac{i}{2}a_{2}\sigma_{3}+\sum_{l=1}^{\infty}a_{l,t}z^{-l}}\right.\\
&\left.{-\frac{i}{2}\sum_{l=1}^{\infty}a_{l+2}\sigma_{3}z^{-l}+\frac{i}{2}q_{0}^{4}(\frac{\sigma_{3}}{z^{2}}+\sum_{l=3}^{\infty}a_{l-2}\sigma_{3}z^{-l})}\right)e^{i\theta(z)\sigma_{3}},
\end{aligned}
\end{equation}
\begin{equation}\label{b41}
\begin{aligned}
-2ik^{2}\sigma_{3}\psi=\left({-\frac{i}{2}\sigma_{3}z^{2}-\frac{i}{2}\sigma_{3}a_{1}z-i\nu q_{0}^{2}\sigma_{3}-\frac{i}{2}\sigma_{3}a_{2}-\frac{i}{2}\sigma_{3}\sum_{l=1}^{\infty}a_{l+2}z^{-l}}\right.\\ \left.{-i\nu q_{0}^{2}\sigma_{3}\sum_{l=1}^{\infty}a_{l}z^{-l}-\frac{i}{2}q_{0}^{4}\sigma_{3}(\frac{1}{z^{2}}+\sum_{l=3}^{\infty}a_{l-2}z^{-l})}\right)e^{i\theta(z)\sigma_{3}},
\end{aligned}
\end{equation}
\begin{equation}\label{b42}
-2kQ\psi=\left(-Qz-Qa_{1}-\sum_{l=1}^{\infty}Qa_{l+1}z^{-l}-\nu q_{0}^{2}(\frac{Q}{z}+\sum_{l=2}^{\infty}Qa_{l-1}z^{-l})\right)e^{i\theta(z)\sigma_{3}},
\end{equation}
\begin{equation}\label{b43}
i\sigma_{3}(Q_{x}-Q^{2}+\nu q_{0}^{2})\psi=\left(i\sigma_{3}(Q_{x}-Q^{2}-q_{0}^{2})+\sum_{l=1}^{\infty}i\sigma_{3}(Q_{x}-Q^{2}-q_{0}^{2})a_{l}z^{-l}\right)e^{i\theta(z)\sigma_{3}},
\end{equation}
which imply that
\begin{equation}\label{b44}
\psi_t=\left(-\frac{i}{2}\sigma_{3}z^{2}-\frac{i}{2}a_{1}\sigma_{3}z-\frac{i}{2}a_{2}\sigma_{3}+O(\frac{1}{z})\right)e^{i\theta(z)\sigma_{3}}, \quad z\rightarrow\infty,
\end{equation}
and
\begin{equation}\label{b45}
\begin{aligned}
&-2ik^{2}\sigma_{3}\psi-2kQ\psi+i\sigma_{3}(Q_{x}-Q^{2}+\nu q_{0}^{2})\psi\\
&\quad=\left(-\frac{i}{2}\sigma_{3}z^{2}-\frac{i}{2}\sigma_{3}a_{1}z-Qz-Qa_{1}-i\nu q_{0}^{2}\sigma_{3}\right.\\
&\quad\qquad\left.-\frac{i}{2}\sigma_{3}a_{2}+i\sigma_{3}(Q_{x}-Q^{2}-q_{0}^{2})+O(\frac{1}{z})\right)e^{i\theta(z)\sigma_{3}}, \quad z\rightarrow\infty.
\end{aligned}
\end{equation}
Using (\ref{c15})-(\ref{c17}), equation (\ref{b45}) can be further reduced to
\begin{equation}\label{b47}
\begin{aligned}
&-2ik^{2}\sigma_{3}\psi-2kQ\psi+i\sigma_{3}(Q_{x}-Q^{2}+\nu q_{0}^{2})\psi\\
&\quad =\left(-\frac{i}{2}\sigma_{3}z^{2}-\frac{i}{2}a_{1}\sigma_{3}z-\frac{i}{2}a_{2}\sigma_{3}+O(\frac{1}{z})\right)e^{i\theta(z)\sigma_{3}}, \quad z\rightarrow\infty.
\end{aligned}
\end{equation}
Equations (\ref{b44}) and (\ref{b47}) imply that the Laurent series of $\partial_{t}\psi$ and $-2ik^{2}\sigma_{3}\psi-2kQ\psi+i\sigma_{3}(Q_{x}-Q^{2}+\nu q_{0}^{2})\psi$ at $z\to\infty$ share the same principal part.

Using (\ref{b20}), (\ref{c14}) and (\ref{c15}), we find from (\ref{b15}) that
\begin{equation}\label{b52}
\psi_t=\left(\frac{i}{2}q_{0}^{4}(b_{1}\sigma_{3}z^{-1}+b_{0}\sigma_{3}z^{-2}+b_{-1}\sigma_{3}z^{-3})+O(1)\right)e^{i\theta(z)\sigma_{3}}, \quad z\rightarrow0,
\end{equation}
and
\begin{equation}\label{b55}
\begin{aligned}
&-2ik^{2}\sigma_{3}\psi-2kQ\psi+i\sigma_{3}(Q_{x}-Q^{2}+\nu q_{0}^{2})\psi\\
&=\left(\frac{i}{2}q_{0}^{4}(b_{1}\sigma_{3}z^{-1}+b_{0}\sigma_{3}z^{-2}+b_{-1}\sigma_{3}z^{-3})+O(1)\right)e^{i\theta(z)\sigma_{3}}, \quad z\rightarrow0.
\end{aligned}
\end{equation}

Since $\partial_{t}\psi\in\mathcal{F}$ and has the asymptotic behaviors given by (\ref{b44}) and (\ref{b52}), then
\begin{equation}\label{b56}
\begin{aligned}
\mathcal{N}(\psi_t)=&-\frac{i}{2}\sigma_{3}z^{2}-\frac{i}{2}a_{1}\sigma_{3}z-\frac{i}{2}a_{2}\sigma_{3}\\
&+\frac{i}{2}q_{0}^{4}(b_{1}\sigma_{3}z^{-1}+b_{0}\sigma_{3}z^{-2}+b_{-1}\sigma_{3}z^{-3}).
\end{aligned}
\end{equation}
From (\ref{b47}) and (\ref{b55}), we find that
\begin{equation}\label{c26}
\mathcal{N}(\psi_t)=\mathcal{N}\left(-2ik^{2}\sigma_{3}\psi-2kQ\psi+i\sigma_{3}(Q_{x}-Q^{2}+\nu q_{0}^{2})\psi\right).
\end{equation}
From this equation and the fact that $-2ik^{2}\sigma_{3}\psi-2kQ\psi+i\sigma_{3}(Q_{x}-Q^{2}-q_{0}^{2})\psi\in\mathcal{F}$, we obtain the temporal linear spectral problem
\begin{equation}\label{b57}
\psi_t=-2ik^{2}\sigma_{3}\psi-2kQ\psi+i\sigma_{3}(Q_{x}-Q^{2}+\nu q_{0}^{2})\psi,
\end{equation}
in view of identity (\ref{c1}).

To derive the focusing/defocusing NLS equation (\ref{a1}) with NZBC (\ref{a2}), we need to introduce the symmetry condition on the off-diagonal matrix $Q$ in (\ref{c15}) as
\begin{equation}\label{c28}
 \sigma_\nu\bar{Q}\sigma_\nu =Q, \quad  \sigma_\nu=\left\{\begin{matrix}
 \sigma_2,&\nu=-1,\\
\sigma_1,& \nu=1,
\end{matrix}
 \right.,
\end{equation}
which implies that the potential $Q$ takes the following form
\begin{equation}\label{c29}
 Q=\left(\begin{matrix}
 0&q\\
\nu\bar{q}&0
\end{matrix}
 \right),
\end{equation}
and $\sigma_\nu\overline{U(x,t;\bar{z})}\sigma_\nu=U(z;x,t)$, where $U(z;x,t)=ik(z)\sigma_3+Q$.
For the linear system (\ref{c15}) and (\ref{b57}) with the boundary condition (\ref{b13}),
in addition to the first symmetry condition (\ref{b13a}), the matrix eigenfunction $\psi(z;x,t)$ admits another symmetry condition
\begin{equation}\label{c30}
 \psi(z;x,t)=\sigma_\nu\overline{\psi(x,t;\bar{z})}\sigma_\nu.
\end{equation}

It is remarked that the first symmetry condition (\ref{b13a}) plays the role of construction of the linear spectral problems (\ref{c15}) and (\ref{b57}), and the second symmetry condition (\ref{c30}) is about to curb the potential matrix $Q$. As a result, the focusing NLS equation (\ref{a1}), is equivalent to the compatibility condition of the linear system (\ref{c15}) and (\ref{b57}) with $Q$ given by (\ref{c29}).

\section{NLS equation and conservation laws}
In this section, we consider that the $2\times2$ distribution $R(z;x,t)$ admits the following properties

(i) The matrix $R(z;x,t)$ has zero diagonal part;

(ii) The time evolution of $R(z;x,t)$ is $\partial_tR(z;x,t)=p(z;x,t)\sigma_3R(z;x,t)$, where $p$ is a scaler function $$p(z;x,t)=2i\partial_t\theta(z)=-4i\lambda(z)k(z).$$
Here $\lambda(z), k(z)$ and $\theta(z)=\theta(z;x,t)$ are defined in (\ref{a7}).\\

Given the distribution, we can define a new Dbar problem
\begin{equation}\label{e1}
 \bar\partial\hat\chi(z;x,t)=\hat\chi(z;x,t)R(z;x,t), \quad z\in\mathbb{C}^0.
\end{equation}
and denote the associated solution space by $\mathcal{\hat{F}}$. It is verified that if $\hat\chi\in \mathcal{\hat{F}}$, then
\begin{equation}\label{e2}
 V(z;x,t)\hat\chi\in \mathcal{\hat{F}}, \quad \hat\chi_t+\frac{1}{2}p(z;x,t)\hat\chi\sigma_3\in\mathcal{\hat{F}},
\end{equation}
for some matrix $V(z;x,t)=\sum_{n=-M}^NV_n(x,t)z^n$, where $N,M\in\mathbb{N}$. In fact, the first property is obviously. To prove the second property, we let $\hat{X}(z;x,t)=\partial_t\hat\chi+\frac{1}{2}p(z;x,t)\hat\chi\sigma_3$, then, using the properties (i) and (ii), we find for $z\in\mathbb{C}^0$
$$\begin{aligned}
\bar\partial\hat{X}=&(\hat\chi_t)R+p\hat\chi\sigma_3 R+\frac{1}{2}p\hat\chi R\sigma_3\\
=&(\hat\chi_t)R+p\hat\chi\sigma_3 R-\frac{1}{2}p\hat\chi\sigma_3R\\
=&(\hat\chi_t+\frac{1}{2}p\hat\chi\sigma_3)R=\hat{X}R.
\end{aligned}$$

In particular, if
$$R(z;x,t)=e^{i\theta(z)\sigma_3}r(z)e^{-i\theta(z)\sigma_3},$$
and the distribution $r(z)$ in (\ref{b1}) admit the property (i), then the matrix function $\hat\psi(z;x,t)$ is a solution of the Dbar problem (\ref{e1}) and takes the following asymptotic behavior
\begin{equation}\label{e3}
\begin{aligned}
\hat\psi(z;x,t)=\sum\limits_{l=0}^\infty a_l(x,t)z^{-l}, \quad z\to\infty,\\
\hat\psi(z;x,t)=\sum\limits_{m=-1}^\infty b_m(x,t)z^{m}, \quad z\to0,
\end{aligned}
\end{equation}
where $a_l(x,t)$ and $b_m(x,t)$ are given in (\ref{b16}) and (\ref{b17}), respectively.
By the properties (\ref{e2}), we know that there is a certain matrix $T(z;x,t)$ admitting the equation
\begin{equation}\label{e4}
\hat\psi_t(z;x,t)+\frac{1}{2}p(z;x,t)\hat\psi(z,x,t)\sigma_3=T(z;x,t)\hat\psi(z;x,t).
\end{equation}
In fact, the matrix $T(z;x,t)$ can be found in the following form
\begin{equation}\label{e5}
 T(z;x,t)=-2ik^{2}\sigma_{3}-2kQ+i\sigma_{3}(Q_{x}-Q^{2}+\nu q_{0}^{2}),
\end{equation}
in view of the temporal linear spectral problem (\ref{b57}).

We note that the trace of distribution $R(z;x,t)$ is zero, then from the Dbar problem (\ref{e1})
\begin{equation}\label{e6}
 \bar\partial\det\hat\psi(z;x,t)=0, \quad z\in\mathbb{C}^0,
\end{equation}
which implies that $\det\hat\psi(z;x,t)$ is analytic in $\mathbb{C}^0$. Then using the Cauchy integral formula and the asymptotic behaviors (\ref{e3}), we find
\begin{equation}\label{e7}
 \det\hat\psi(z;x,t)=1-\nu\frac{q_0^2}{z^2}:=\gamma.
\end{equation}
In fact, by the Cauchy integral formula the asymptotic behaviors (\ref{e3}), we know that
$$\begin{aligned} \det\hat\psi(z;x,t)=&\frac{1}{2\pi i}\int_{\Gamma_R+\Gamma_\varepsilon^-}\frac{\det\hat\psi(\mu;x,t)}{\mu-z}d\mu\\
=&\lim\limits_{R\to\infty}\frac{1}{2\pi i}\int_{\Gamma_R}\frac{1}{\mu-z}d\mu
+\nu\lim\limits_{\varepsilon\to0}\frac{1}{2\pi i}\int_{\Gamma_\varepsilon}\frac{q_0^2/(\mu-z)}{\mu^2}d\mu\\
=&1-\nu\frac{q_0^2}{z^2}.
\end{aligned}$$

For $z\neq\pm iq_0, (\nu=-1)$ or $z\neq\pm q_0, (\nu=1)$, we have $\gamma\hat\psi^{-1}=\sigma_2\hat\psi^T\sigma_2$. Thus, equation (\ref{e4}) can be rewritten as
\begin{equation}\label{e8}
 \gamma T(z;x,t)\sigma_2=(\hat\psi_t)(z;x,t)\sigma_2\hat\psi^T(z;x,t)-\frac{i}{2}p(z;x,t)\Psi(z;x,t),
\end{equation}
where $\Psi(z;x,t)=\hat\psi(z;x,t)\sigma_1\hat\psi^T(z;x,t)=\Psi^T(z;x,t)$. Note that $\Psi(z;x,t)$ admits the following asymptotic behavior as $z\to\infty$
\begin{equation}\label{e9}
 \Psi(z;x,t)=\sum\limits_{n=0}^\infty \frac{\Psi_n(x,t)}{z^n}, \quad z\to\infty,
\end{equation}
where
\begin{equation}\label{e10}
 \begin{aligned}
 \Psi_0=\sigma_1, \quad \Psi_n(x,t)=\sum\limits_{m=0}^na_m(x,t)\sigma_1a_{n-m}^T(x,t), \quad (n\geq1).
 \end{aligned}
\end{equation}
Similarly, at $z=0$, we have
\begin{equation}\label{e11}
 \Psi(z;x,t)=\sum\limits_{n=-2}^\infty\tilde\Psi_n(x,t)z^n, \quad z\to0,
\end{equation}
where
\begin{equation}\label{e12}
 \tilde\Psi_{-2}=\nu q_0^2\sigma_1, \quad \tilde\Psi_n=\sum\limits_{m=-1}^{n+1}b_m\sigma_1b_{n-m}^T, \quad (n\geq-1).
\end{equation}

Substituting the expansions (\ref{e4}) and (\ref{e9}) at $z\to\infty$ into (\ref{e8}), and considering the $O(z^{-1})$ items, we have
\begin{equation}\label{e13}
 a_{1,t}(x,t)=\frac{1}{2}\Psi_3(x,t)\sigma_2.
\end{equation}
Since $a_1(x,t)$ can be regarded as the function of the potential matrix $Q$ by (\ref{c15}) and (\ref{c17}), so $\Psi_3(x,t)$ should also be the function of $Q$. Next, we will find the expression of $\Psi_3(x,t)$ in terms of $Q$.

Note that $\hat\psi(z;x,t)$ admits the linear problem
\begin{equation}\label{e14}
\hat\psi_x(z;x,t)=-i\lambda(z)\hat\psi(z;x,t)\sigma_3+ik(z)\sigma_3\hat\psi(z;x,t)+Q\hat\psi(z;x,,t),
\end{equation}
in view of the spatial linear spectral problem (\ref{c15}). Hence, $\Psi(z;x,t)$ satisfies the following linear problem
\begin{equation}\label{e15}
\Psi_x(z;x,t)=ik(z)(\sigma_3\Psi(z;x,t)+\Psi(z;x,t)\sigma_3)+Q\Psi(z;x,t)+\Psi(z;x,t)Q^T,
\end{equation}
which can be rewritten as
\begin{equation}\label{e16}
\Psi_x^{[d]}(z;x,t)=2ik(z)\sigma_3\Psi^{[d]}(z;x,t)+Q\Psi^{[o]}(z;x,t)+\Psi^{[o]}(z;x,t)Q^T,
\end{equation}
and
\begin{equation}\label{e17}
\Psi_x^{[o]}(z;x,t)=Q\Psi^{[d]}(z;x,t)+\Psi^{[d]}(z;x,t)Q^T,
\end{equation}
where $\Psi^{[d]}$ and $\Psi^{[o]}$ denote the diagonal part and off-diagonal part of the matrix $\Psi$.
Substituting the expansion (\ref{e9}) at $z\to\infty$ into (\ref{e16}) and (\ref{e17}), we find
\begin{equation}\label{e18}
 \begin{aligned}
&i\sigma_3\Psi_1^{[d]}+2Q\sigma_1=0,\\
&i\sigma_3\Psi_2^{[d]}=\Psi_{1,x}^{[d]}-(Q\Psi_1^{[o]}+\Psi_1^{[o]}Q^T),\\
&i\sigma_3\Psi_{n+1}^{[d]}=\Psi_{n,x}^{[d]}+iq_0^2\sigma_3\Psi_{n-1}^{[d]}-(Q\Psi_n^{[o]}+\Psi_n^{[o]}Q^T), \quad n=2,3,\cdots,
 \end{aligned}
\end{equation}
and
\begin{equation}\label{e19}
\Psi_{n,x}^{[o]}=Q\Psi_n^{[d]}+\Psi_n^{[d]}Q^T, \quad n=1,2,3,\cdots.
\end{equation}
Here and after the variables $(x,t)$ are omitted for simplicity. Equations (\ref{e18}) and (\ref{e19}) are recurrent formula, from which we can find all $\Psi_n$, for example
\begin{equation}\label{e20}
 \begin{aligned}
&\Psi_1^{[d]}=2Q\sigma_2, \quad \Psi_1^{[o]}=0;\\
&\Psi_2^{[d]}=2Q_x\sigma_1, \quad \Psi_2^{[o]}=(2Q^2-\nu q_0^2)\sigma_1;\\
&\Psi_3^{[d]}=-2Q_{xx}\sigma_2+4(Q^2-\nu q_0^2)Q\sigma_2, \quad \Psi_3^{[o]} =2(Q_xQ-QQ_x)\sigma_2;\\
&\Psi_4^{[d]}=(-2Q_{xxx}+12Q^2Q_x-6\nu q_0^2Q_x)\sigma_2,\\
&\Psi_4^{[o]}=-2(QQ_{xx}+Q_{xx}Q-Q_x^2)\sigma_1+6(Q^2-\nu q_0^2)Q^2\sigma_1,\\
&\qquad\cdots \qquad \cdots \qquad \cdots
 \end{aligned}
\end{equation}

With the expressions of $\Psi_n$ in hand, from the off-diagonal part of equation (\ref{e13}), we find the nonlinear equation
\begin{equation}\label{e21}
i\sigma_3Q_t+Q_{xx}-2(Q^2-\nu q_0^2)Q=0,
\end{equation}
which implies the nonlinear NLS equation (\ref{a1}). In addition, from the diagonal part of (\ref{e13}), we get the first conservation law
\begin{equation}\label{e22}
 i(|q|^2-q_0^2)_t=(q\bar{q}_x-q_x\bar{q})_x.
\end{equation}

Similarly,  the $O(z^{-2})$ terms in the expansion of (\ref{e8}) by substituting (\ref{e9}) take the following form
\begin{equation}\label{e23}
 a_{2,t}\sigma_2+a_{1,t}\sigma_2a_1^T=\nu q_0^2Q_x\sigma_1+\nu q_0^2(Q^2-\nu q_0^2)\sigma_1+\frac{1}{2}\Psi_4.
\end{equation}
It is noted that
\begin{equation}\label{e25}
\begin{aligned}
&a_1^{[0]}=-iQ\sigma_3, \quad a_{1,x}^{[d]}=-i(Q^2-\nu q_0^2)\sigma_3, \quad a_2^{[0]}=Q_x-Q\partial_x^{-1}(Q^2-\nu q_0^2), \\
& \quad a_{2,x}^{[d]}=QQ_x-(Q^2-\nu q_0^2)\partial_x^{-1}(Q^2-\nu q_0^2),
\end{aligned}
\end{equation}
then the diagonal part of equation (\ref{e23}) also reduces to the nonlinear equation (\ref{e21}), and the off-diagonal part gives the second conservation law
\begin{equation}\label{e24}
 i(q\bar{q}_x)_t=\big[q\bar{q}_{xx}-|q_x|^2-\nu(|q|^2-q_0^2)(|q|^2+q_0^2)\big]_x.
\end{equation}

It is remarked that the NLS equation (\ref{e21}) and conversation law (\ref{e22}) can also be derived from equation (\ref{e8}) by substituting the expansion (\ref{e11}) at $z=0$ in terms of the symmetry condition (\ref{b13a}).
The other conservation laws can be found from the diagonal part of the items $O(z^{-n}), (n=3,4,\cdots)$ in the expansion of (\ref{e8}) at $z\to\infty$ \cite{pla117-62}. It is also noted that the focusing and defocusing NLS equation share the same first conversation law (\ref{e22}), but have different second conversation law (\ref{e24}).

\section{Explicit solutions}
From (\ref{c15}) and (\ref{b16}), we know that
\begin{equation}\label{d1}
Q=Q_{0}+\frac{1}{4\pi}\left[\sigma_{3},\iint_{\mathbb{C}^0}\hat{\psi}(z)e^{i\theta(z)\sigma_{3}}r(z)e^{-i\theta(z)\sigma_{3}}\mathrm{d}z\wedge \mathrm{d}{\bar{z}}\right].
\end{equation}
where $\hat{\psi}(z)$ defined by (\ref{b4}) satisfies the integral equation (\ref{b8}). We note that the distribution $r(z)$ in ${\mathbb{C}^0}$ admits
\begin{equation}\label{d1a}
 \overline{r(\bar{z})}=\sigma_\nu r(z)\sigma_\nu,
\end{equation}
and
\begin{equation}\label{d1b}
 r(z)=\frac{1}{\bar{z}^2}\sigma_3Q_0r\big(\nu\frac{q_0^2}{z}\big)\sigma_3Q_0,
\end{equation}
in terms of the symmetry condition (\ref{c30}) and (\ref{b13a}).

\subsection{Solution of focusing NLS with NZBC}

To obtain the soliton solutions of the focusing NLS equation ($\nu=-1$) with nonzero boundary condition (\ref{a1}), we choose, according to the symmetries (\ref{d1a}) and (\ref{d1b}), the distribution $r(z)$ as
\begin{equation}\label{d2}
\begin{aligned}
r(z)=\pi\sum_{j=1}^{2N}\left(\begin{matrix}
0&\bar{c}_{j}\delta(z-\bar\zeta_{j})\\
-c_{j}\delta(z-\zeta_{j})&0
\end{matrix}\right),
\end{aligned}
\end{equation}
where $c_{j}\in\mathbb{C}, j=1,2,\cdots,N$ and
\begin{equation}\label{d3}
\zeta_{j}=z_{j},\quad \zeta_{j+N}=-\frac{q_{0}^{2}}{\bar{z}_{j}}, \quad c_{N+j}=-\left(\frac{\rho}{\bar{z_j}}\right)^2\bar{c_j}.
\end{equation}

Substituting (\ref{d2}) into (\ref{d1}) and (\ref{b8}), we find the solution of the focusing NLS equation with NZBC
\begin{equation}\label{d4}
 q=\rho+i\frac{\det M^a}{\det M}, \quad M=I+\Omega\bar{\Omega},
\end{equation}
where
\begin{equation}\label{d5}
\begin{aligned}
\Omega=(\Omega_{nj})_{(2N)\times(2N)}, \quad \Omega_{jn}=\frac{g_j}{\zeta_j-\bar{\zeta_n}}, \quad M^a=\left(\begin{matrix}
0&\bar{g}\\
f^T&M
\end{matrix}\right),\\
 g=(g_1,g_2,\cdots,g_{2N}), \quad f=(f_1,f_2,\cdots,f_{2N}),\\
g_j=c_je^{-2i\theta(\zeta_j;x,t)}, \quad f_n=1+i\rho\sum\limits_{j=1}^{2N}\frac{\Omega_{jn}}{\zeta_j}.
\end{aligned}
\end{equation}
It is remarked that if the eigenvalues $z_j$ admit $|z_j|>q_0$ and $\Im{z_j}>0$, then the solution is in correspondence with that obtained by Riemann-Hilbert problem, that is to say, $\rho$ is the boundary condition at $x\to-\infty$. Some special solutions can be obtained by choosing different parameters $N$ and $\zeta_j$, see reference \cite{jmp55-031506}.

\subsection{Solution of defocusing NLS with NZBC}
Next, we consider the explicit solution of the defocusing NLS equation with NZBC. Since the self-adjointness of the linear spectral problem (\ref{c15}) and (\ref{b57}) for the defocusing NLS equation, the associated eigenvalues locate on the circle $|z|=q_0^2$. Hence, to obtain the soliton solution of defocusing NLS equation ($\nu=1$), we take the distribution in the following form
\begin{equation}\label{d6}
\begin{aligned}
r(z)=\pi\sum_{j=1}^{N}\left(\begin{matrix}
0&\bar{d}_{j}\delta(z-\bar\eta_{j})\\
d_{j}\delta(z-\eta_{j})&0
\end{matrix}\right),
\end{aligned}
\end{equation}
where $|\eta_{j}|=q_0$.

It is remarked that the focusing NLS equation with NZBC has two sets of eigenvalues in view of the symmetry conditions  (\ref{d1a}) and (\ref{d1b}) with $\nu=-1$, while for defocusing NLS equation with NZBC, the symmetry conditions  (\ref{d1a}) and (\ref{d1b}) with $\nu=1$ only reduce to one set of eigenvalues.

A similar procedure implies that the solution of the defocusing NLS equation with NZBC can be reconstructed as
\cite{F-T1987,HNN1996,mplb27-1350216}
\begin{equation}\label{d10}
q(x)=\rho+i\frac{\det D^a}{\det D}, \quad D=I-i\rho{G},
\end{equation}
where
\begin{equation}\label{d11}
 \begin{aligned}
&G=(G_{jl})_{N\times N}, \quad G_{jl}=\frac{h_l}{(\bar\eta_j-\eta_l)\eta_l}, \quad D^a=\left(\begin{matrix}
0&\bar{h}\\
E^T&D
\end{matrix}\right),\\
& h=(h_1,\cdots,h_N), \quad E=(1,\cdots,1)_{1\times N}, \quad h_j=d_j e^{-2i\theta(\eta_j)} .
\end{aligned}
\end{equation}

Let $\eta_j=q_0e^{i\alpha_j}, 0<\alpha_j<\pi$, and $d_j$ admit $\arg d_j=\alpha_j-\arg\rho$ and
\begin{equation}\label{d12}
\frac{i\rho d_j}{\eta_j(\eta_j-\bar\eta_j)}=\frac{|d_j|}{2q_0\sin\alpha_j}=e^{2\epsilon_j}, \quad \epsilon_j\in \mathbb{R},
\end{equation}
then
\begin{equation}\label{d13}
 \frac{i\rho h_j}{\eta_j(\eta_j-\bar\eta_j)}=e^{2\vartheta_j}, \quad \vartheta_j=q_0\sin\alpha_j(x-2q_0t\cos\alpha_j)+\epsilon_j.
\end{equation}
In this case, $\rho$ is correspondence with the boundary condition at $x\to+\infty$.

In particular, for $N=1$, we obtain the dark one-soliton \cite{sam131-1}
\begin{equation}\label{d14}
 q(x,t)=\rho e^{-i\alpha_1}[\cos\alpha_1-i\sin\alpha_1\tanh\vartheta_1].
\end{equation}

For $N=2$, the two-soliton solution of the defocusing NLS equation (\ref{a1}) takes the following form
\begin{equation}\label{d15a}
q(x,t)=\rho\left[\frac{1+e^{2\hat\vartheta_1}+e^{2\hat\vartheta_2}+ne^{2(\hat\vartheta_1+\hat\vartheta_2)}}
{1+e^{2\vartheta_1}+e^{2\vartheta_2}+ne^{2(\vartheta_1+\vartheta_2)}}\right],
\end{equation}
where
\begin{equation}\label{d16a}
\begin{aligned}
 \hat\vartheta_j=&\vartheta_j-i\alpha_j, \quad j=1,2,\\
 n=&\frac{1-\cos(\alpha_1-\alpha_2)}{1-\cos(\alpha_1+\alpha_2)}.
\end{aligned}
\end{equation}

Since $0<\alpha_j<\pi, (j=1,2)$ and $\cos(\alpha_1-\alpha_2)-\cos(\alpha_1+\alpha_2)=2\sin\alpha_1\sin\alpha_2>0$, then $0<n<1$. We note from (\ref{d15a}) that
\begin{equation}\label{d16}
 q\sim\left\{\begin{array}{cc}
 \rho,& x\to-\infty,\\
 \rho e^{-2i(\alpha_1+\alpha_2)}, &x\to+\infty.
 \end{array}\right.
\end{equation}
Since $\alpha_1$ and $\alpha_2$ are symmetry in (\ref{d15a}), for the convenience of discussion, we assume $\alpha_1<\alpha_2$ and let $v_j=2q_0\cos\alpha_j, (j=1,2)$, then $v_1>v_2$. Let $X_j=v_jt+x_{j}, \epsilon_j=-(q_0\sin\alpha_j)x_j$, and define the compact domain $\Omega_j$ containing the point $x=X_j$.

As $t\to-\infty$, the domains $\Omega_j$ will separated, and $\Omega_1$ is located on the left of $\Omega_2$.
Now, in the domain $\Omega_1$, $e^{2\vartheta_1}$ and $e^{2\hat\vartheta_1}$ are bounded, and $e^{2\vartheta_2}$ and $e^{2\hat\vartheta_2}$ are $O(e^{(v_1-v_2)t})$. Thus, in $\Omega_1$, we find a dark one-soliton (denoted by $S_1^-$)
\begin{equation}\label{d17}
 \rho e^{-i\alpha_1}\left[\cos\alpha_1-i\sin\alpha_1\tanh\vartheta_1\right].
\end{equation}
In $\Omega_2$, we find another dark one-soliton (denoted by $S_2^-$)
\begin{equation}\label{d18}
 \rho e^{-i(\alpha2+2\alpha_1)}\left[\cos\alpha_2-i\sin\alpha_2\tanh(\vartheta_2-\tau)\right].
\end{equation}
where $\tau=-\ln(n)/2$.

In first case, $\alpha_1<\alpha_2<\pi/2$, or $v_1>v_2>0$, then $S_1^-$ and $S_2^-$ will move to right, and $S_1^-$ will catch up with $S_2^-$ as $t\to0$. After collision, $S_1^-$ will surpass the latter. In second case, $\alpha_1<\pi/2<\alpha_2$, or $v_1>0>v_2$, then $S_1^-$ moves to right and $S_2^-$ moves to left, and they will collide as $t\to0$. In third case, $\pi<\alpha_1<\alpha_2$, or $v_1<v_2<0$, $S_2^-$ and $S_1^-$ will move to left, and $S_1^-$ will also catch up with $S_2^-$ as $t\to0$.

Similarly, as $t\to+\infty$,  $\Omega_1$ is located on the right of $\Omega_2$. We also obtain two dark one-solitons
\begin{equation}\label{d19}
\begin{aligned}
S_1^+:&\quad \rho e^{-i(\alpha_1+2\alpha_2)}\left[\cos\alpha_1-i\sin\alpha_1\tanh(\vartheta_1-\tau)\right], \quad x\in\Omega_1,\\
S_2^+:&\quad \rho e^{-i\alpha_2}\left[\cos\alpha_2-i\sin\alpha_2\tanh\vartheta_2\right], \qquad\qquad \quad x\in\Omega_2.
\end{aligned}
\end{equation}
We note that wave heights of two solitons are $q_0|\cos\alpha_j|, j=1,2$.

As a consequence of the interaction, the phase shift of dark one-soliton in $\Omega_1$ is $-\tau$ and that of $\Omega_2$ is $\tau$. For $N$ soliton, the interaction can be discussed similarly \cite{HNN1996}. We note that there is an energy superposition as collision in second case, and no energy superposition for the first case and the third one.
The plots of the interaction on above three cases  are shown in Figure 1. In addition, the wave height at the collision point, that is, $|q(0,0)|$ (with $q_0=1, \epsilon_j=0$) about $\alpha_1$ and $\alpha_2$ is shown in Figure 2. The value of $|q(0,0)|$ in Figure 2 and $q_0|\cos\alpha_j|$ imply the energy superposition in different cases.
\begin{figure}[h]
  \centering
  \includegraphics[width=4cm,height=4cm]{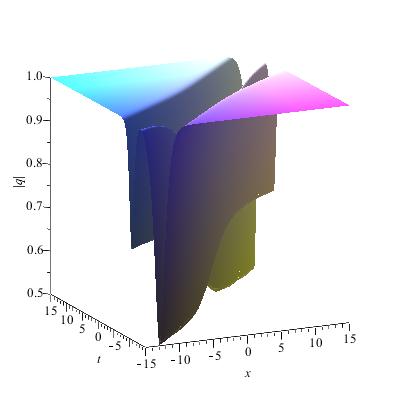} \quad
  \includegraphics[width=4cm,height=4cm]{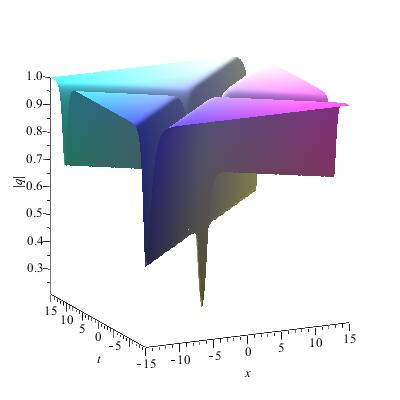} \quad
  \includegraphics[width=4cm,height=4cm]{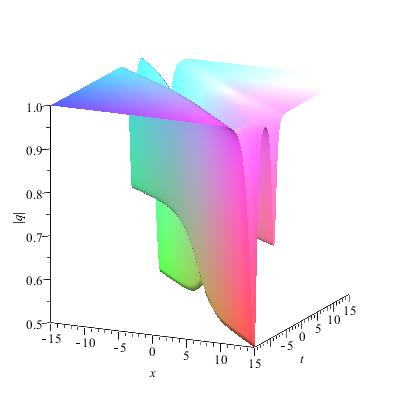}
  \caption{Plots of solution (\ref{d15a}) with $\rho=e^{i\frac{\pi}{4}}, q_0=1, \epsilon_j=0$. In addtion, (left)$\alpha_1=\frac{\pi}{4}, \alpha_2=\frac{\pi}{3}$, (center)$\alpha_1=\frac{\pi}{3}, \alpha_2=\frac{3\pi}{4}$ and (right)$\alpha_1=\frac{2\pi}{3}, \alpha_2=\frac{3\pi}{4}$ .}
\end{figure}

\begin{figure}[h]
  \centering
  \includegraphics[width=4.5cm,height=4.5cm]{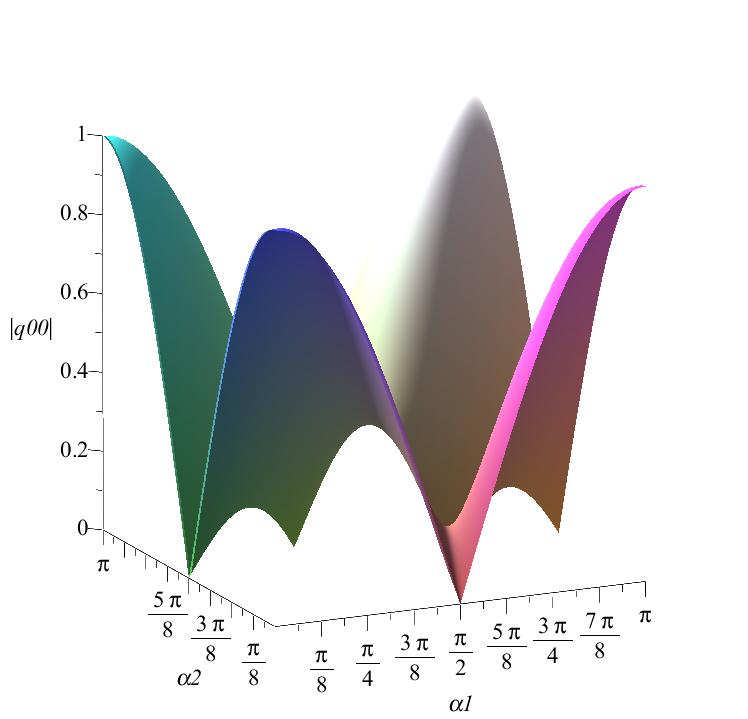} \quad
   \includegraphics[width=4cm,height=4cm]{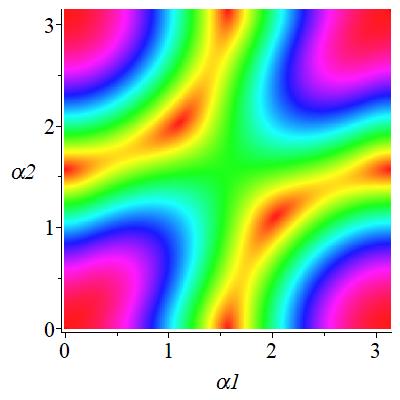}
  \caption{The wave height of $|q(0,0)|$ in solution (\ref{d15a}) with $q_0=1,\epsilon_j=0$.}
\end{figure}

Furthermore, We find that the two wave trains lie in lines in $x$-$t$ plane with slopes $\kappa_j=1/v_j=(2q_0\cos\alpha_j)^{-1}$ which imply that the "collision angle"
denoted by $\Delta\alpha$ admits
\begin{equation}\label{d21}
\tan\Delta\alpha=\frac{2q_0(\cos\alpha_1-\cos\alpha_2)}
{1+4q_0^2\cos\alpha_1\cos\alpha_2}:=F(q_0).
\end{equation}
If the two eigenvalues $\eta_j=q_0e^{i\alpha_j}$ or $\alpha_j, (0<\alpha_1<\alpha_2<\pi)$ are fixed, then the angle is determined by the initial amplitude $q_0$. In fact,
\begin{equation}\label{d24}
 F'(q_0)=\frac{2(\cos\alpha_1-\cos\alpha_2)(1-4q_0^2\cos\alpha_1\cos\alpha_2)}{(1+4q_0^2\cos\alpha_1\cos\alpha_2)^2}.
\end{equation}

For $\cos\alpha_1\cos\alpha_2>0$, the angle $\Delta\alpha$ in (\ref{d21}) takes the maximum
$$\arctan\left(\frac{\cos\alpha_1-\cos\alpha_2}{2\sqrt{\cos\alpha_1\cos\alpha_2}}\right),$$
at $q_0=\tilde{q}_0$, where
$$\tilde{q}_0=\frac{1}{2\sqrt{\cos(\alpha_1)\cos(\alpha2)}}.$$
We called collision in this case as the glancing-crossing collision. We note that in the glancing-crossing collision, there is no energy superposition. See Figure 3.
\begin{figure}[h]
  \centering
  \includegraphics[width=4.5cm,height=4.5cm]{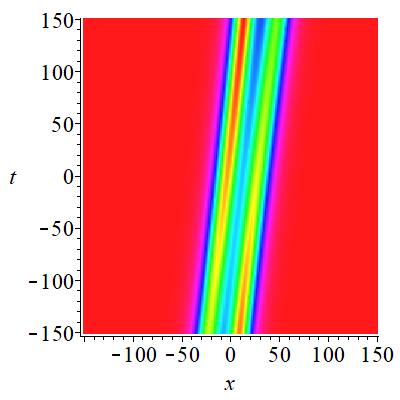} \quad
  \includegraphics[width=4.5cm,height=4.5cm]{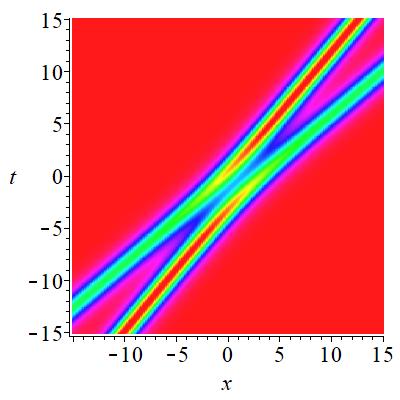} \quad
  \includegraphics[width=4.5cm,height=4.5cm]{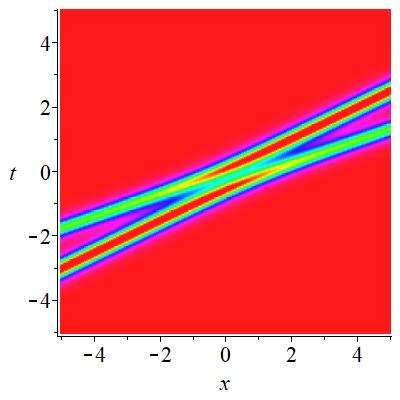}
  \caption{Solution (\ref{d15a}) with $\rho=q_0e^{i\frac{\pi}{4}}, \epsilon_j=0, \alpha_1=\frac{\pi}{4}, \alpha_2=\frac{\pi}{3}$. In addition, (left) with $q_0=0.1$; (center) with $q_0=\frac{1}{2\sqrt{\cos(\alpha_1)\cos(\alpha_2)}}$, and (right) with $q_0=2$.
  }
\end{figure}

For $\cos\alpha_1\cos\alpha_2<0$, the angle $\Delta\alpha$ in (\ref{d21}) monotonically increases from 0 to $\pi$ with respect to $q_0$ changing from 0 to $\infty$, and takes $\pi/2$ at the point $$q_0^*=\frac{1}{2\sqrt{-\cos(\alpha_1)\cos(\alpha2)}}.$$
For example, for the eigenvalues in Figure 1 (center), we can change the initial amplitude $q_0$ to modify the collision angle.
The energy superposition still exists despite the glancing collision, see Figure 4 ($q_0=0.1, q_0=1/(2\sqrt{-\cos(\alpha_1)\cos(\alpha_2)}$ and $q_0=2$).
\begin{figure}[h]
  \centering
  \includegraphics[width=4.5cm,height=4.5cm]{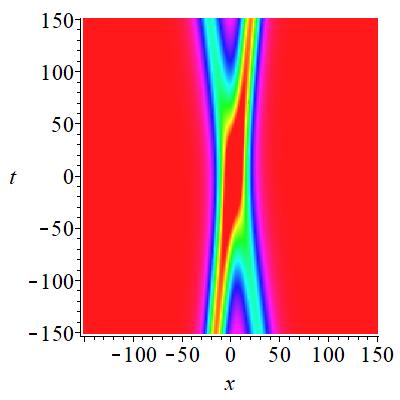}  \quad
  \includegraphics[width=4.5cm,height=4.5cm]{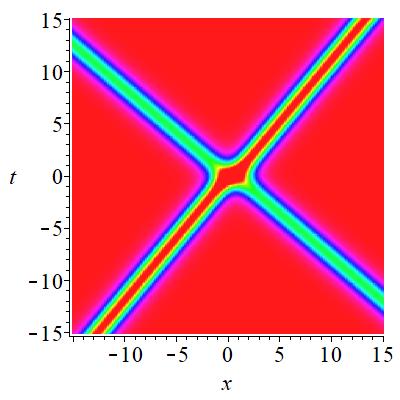} \quad
  \includegraphics[width=4.5cm,height=4.5cm]{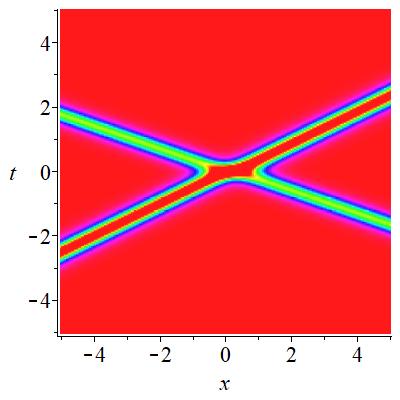}
    \caption{Solution (\ref{d15a})  with $\rho=q_0e^{i\frac{\pi}{4}}, \epsilon_j=0, \alpha_1=\frac{\pi}{3}, \alpha_2=\frac{3\pi}{4}$. In addition, (left) with $q_0=0.1$; (center) with $q_0=\frac{1}{2\sqrt{-\cos(\alpha_1)\cos(\alpha_2)}}$, and (right) with $q_0=2$.}
\end{figure}

In addition, the direction of the wave trains is also determined by $q_0$, that is, the two wave trains is vertical in $x$-$t$ plane as $q_0\to0$ and is horizontal as $q_0\to\infty$.

Next, for $N=2$, we consider a particular case: $\eta_2=-\bar\eta_1$, that is $\alpha_1+\alpha_2=\pi, 0<\alpha_1<\pi/2$, then two-soliton (\ref{d15a}) reduces to
\begin{equation}\label{d22}
 q(x,t)=\rho\left[\frac{\cos(\alpha_1)\cosh(X)+\cos(2\alpha_1)\cosh(T)+i\sin(2\alpha_1)\sinh(T)}
 {\cos(\alpha_1)\cosh(X)+\cosh(T)}\right],
\end{equation}
where
\begin{equation}\label{d23}
 \begin{aligned}
X=&2q_0\sin(\alpha_1)\cdot x+\ln(\cos(\alpha_1))+\epsilon_1+\epsilon_2,\\
T=&2q_0^2\sin(2\alpha_1)\cdot t+\epsilon_2-\epsilon_1.
 \end{aligned}
\end{equation}
The solution is a special case of $\cos\alpha_1\cos\alpha_2<0$.

\section{Conclusions}
In this paper, we considered the Dbar-problem with non-canonical normalization condition, which was equivalent to an inhomogeneous integral equation. It is assumed that the associated homogeneous integral equation only has zero solution. Since the solution of the Dbar-problem was meromorphic outside an annulus with center 0 and satisfied a local Dbar problem inside the annulus. We developed the relevant theory to find new solutions of the Dbar problem with different inhomogeneous terms.

We extended the Dbar dressing method to study the focusing and defocusing NLS equation with nonzero boundary condition. To establish the relation between the NLS potential and the Dbar data, we constructed the linear system associated with the NLS equation. To obtain the focusing and defocusing NLS equation, we introduced different symmetry conditions, which implied different forms of the distribution or the spectral transform matrix. For focusing NLS equation with nonzero boundary condition, $2N$ discrete eigenvalues were introduced in the distribution, and $N$-soliton solution was given.

For defocusing NLS equation, the associated linear spectral problem is self-adjoint, which implied that the eigenvalues located on the circle $|z|=q_0$. One-soliton, two-soliton and $N$-soliton of the defocusing NLS equation with nonzero boundary condition were given. The collision angle of dark two-soltion was found to be related to the boundary condition. %

A special distribution and a symmetry matrix function were introduced to construct the NLS equation( with nonzero bounary condition). The asymptotic expansion of the symmetry matrix was considered to obtain a series of equations. The off-diagonal part of these equations gave the NLS equation, and diagonal part provided the conservation laws.

It is seems that the analytical domains of the eigenfunctions are not important for the Dbar problem. Thus, unlike the Jost functions defined in the Riemann-Hilbert approach, the boundary condition $\rho$ at $x\to+\infty$ or $x\to-\infty$ for the Dbar approach seems not crucial. However, if the special distribution with Dirac delta functions is chosen, the boundary condition can be confined to a certain direction (see (\ref{d16})).

\section*{Acknowledgment}
This work was supported by the National Natural Science Foundation of PR China [Grant number 11471295, 11971442].

\def\thesection{Appendixes \Alph{section}}
\renewcommand{\thelemma}{A\arabic{lemma}}
\setcounter{equation}{0}
\section*{Appendix}

{\bf Lemma 1}~~ {\sl
Suppose $f(z)$ is a meromorphic function in the domain $D$ and has only one $m$th-order pole $z_0\in D$. Let the circle $\Gamma_\varepsilon=\{z:|z-z_0|=\varepsilon\}$ in $D$, then
\[
\frac{1}{2\pi i}\int_{\Gamma_\varepsilon^-}\frac{f(k)}{k-z}{\rm d} k=\sum\limits_{j=1}^{m}\frac{a_{m-j}}{(z-z_0)^{j}},  \quad |z-z_0|>\varepsilon, \eqno(A.1)
\]
where $\Gamma_\varepsilon^-$ denotes a contour taking the negative direction of the circle $\Gamma_\varepsilon$ and
\[
f(z)=\sum\limits_{j=1}^{m}\frac{a_{m-j}}{(z-z_0)^{j}}+\sum\limits_{l=0}^\infty \tilde{a}_l(z-z_0)^l, \quad 0<|z-z_0|<\varepsilon. \eqno(A.2)
\]
}
{\bf Proof}. Suppose $C$ is a simple closed contour in $D$ and encloses the circle $\Gamma_\varepsilon$. For all $z$ in the region with boundary $C$, and admitting $|z-z_0|>\varepsilon$, we have
\[\frac{1}{2\pi i}\int_C\frac{f(k)}{k-z}{\rm d} k=f(z)+\frac{1}{(m-1)!}\lim\limits_{k\to z_0}\frac{{\rm d}^{m-1}}{{\rm d} k^{m-1}}\frac{f(k)(k-z_0)^m}{k-z}. \eqno(A.3)\]
Since $z_0$ is a $m$th-order pole of $f$, and $0<|k-z_0|<\varepsilon$, then
\[(k-z_0)^mf(k)=\sum\limits_{j=0}^{m-1} a_{j}(k-z_0)^{j}+\sum\limits_{j=m}^\infty \tilde{a}_{j-m}(k-z_0)^{j}.\]
In addition, for $|z-z_0|>\varepsilon$, we have
\[\frac{1}{k-z}=\frac{-1}{z-z_0}\frac{1}{1-\frac{k-z_0}{z-z_0}}
=-\sum\limits_{l=0}^\infty\frac{(k-z_0)^l}{(z-z_0)^{l+1}},\quad 0<|k-z_0|<\varepsilon.\]
Thus,
\[
\frac{1}{(m-1)!}\lim\limits_{k\to z_0}\frac{{\rm d}^{m-1}}{{\rm d} k^{m-1}}\frac{f(k)(k-z_0)^m}{k-z}
=-\sum\limits_{j=1}^{m}\frac{a_{m-j}}{(z-z_0)^{j}}. \eqno(A.4)
\]
Substituting (A.4) into (A.3) implies that
\[\frac{1}{2\pi i}\int_C\frac{f(k)}{k-z}{\rm d} k=f(z)-\sum\limits_{j=1}^{m}\frac{a_{m-j}}{(z-z_0)^{j}}. \eqno(A.5)\]

On the other hand, for the region with boundary $C$ and $\Gamma_\varepsilon$, we apply the Cauchy's formula and find
\[f(z)=\frac{1}{2\pi i}\int_{C+\Gamma_\varepsilon^-}\frac{f(k)}{k-z}{\rm d} k=\frac{1}{2\pi i}\int_C\frac{f(k)}{k-z}{\rm d} k+\frac{1}{2\pi i}\int_{\Gamma_\varepsilon^-}\frac{f(k)}{k-z}{\rm d} k.\eqno(A.6)\]
The Lemma 1 is proved by comparing equations (A.5) and (A.6). It is noted that
\[\int_{\Gamma_\varepsilon^-}\frac{f(k)}{k-z}{\rm d} k=\lim\limits_{\varepsilon\to 0}\int_{\Gamma_\varepsilon^-}\frac{f(k)}{k-z}{\rm d} k.\]

{\bf Lemma 2}~~{\sl Suppose that $f(z)$ is a meromorphic function in the region $|z|> \epsilon>0$ and satisfies the asymptotic behavior at the infinite point
\[f(z)= \sum\limits_{j=0}^nb_jz^j+O(1/z),\quad z\to\infty, \eqno(A.7)\]
then for the contour $\Gamma_R=\{z:|z|=R>\epsilon\}$
\[\frac{1}{2\pi i}\oint_{\Gamma_R}\frac{f(k)}{k-z }{\rm d}k
=\sum\limits_{j=0}^nb_jz^j, \quad r<|z|<R, \eqno(A.8)\]
for some $r>\epsilon$.
}

{\bf Proof}~~ Consider the contour integral on the circle $\Gamma_r=\{k:|k|=r, \epsilon<r<R\}$
\[\frac{1}{2\pi i}\oint_{\Gamma_r^-}\frac{f(k)}{k-z }{\rm d}k=-\frac{1}{2\pi i}\oint_{\Gamma_r}\frac{f(k)}{k-z }{\rm d}k, \quad r<|z|<R.\]
Suppose $z=\zeta^{-1}$ and make the transformation $k=\mu^{-1}, f(\mu^{-1})=g(\mu)$. Then the circle $\Gamma_r$ is transformed to the circle $C=\{\mu:|\mu|=r^{-1}\}$. Noting the direction of the contours will change after the transformation, we have
\[\frac{1}{2\pi i}\oint_{\Gamma_r^-}\frac{f(k)}{k-z }{\rm d}k=\frac{\zeta}{2\pi i}\oint_C\frac{g(\mu)}{\mu(\mu-\zeta)}{\rm d}\mu. \eqno(A.9)\]
Since $0<|\zeta|<r^{-1}$ and
\[g(\mu)=\sum\limits_{j=0}^n\frac{b_j}{\mu^j}+O(\mu), \quad \mu\to0,\]
then
\[\frac{1}{2\pi i}\oint_C\frac{g(\mu)}{\mu(\mu-\zeta)}{\rm d}\mu=\frac{g(\zeta)}{\zeta}+\frac{1}{n!}\lim\limits_{\mu\to0}\frac{{\rm d}^n}{{\rm d}\mu^n}\big[\frac{g(\mu)\mu^n}{\mu-\zeta}\big]. \eqno(A.10)\]
For fixed $\zeta$, we know that
\[\frac{g(\mu)\mu^n}{\mu-\zeta}=-\sum\limits_{l=0}^n\big(\sum\limits_{j=0}^l\frac{b_{n-l+j}}{\zeta^{j+1}}\big)\mu^l+O(\mu^{n+1}), \quad \mu\to0. \eqno(A.11)\]
Substituting (A.11) into (A.10), we find
\[\frac{1}{2\pi i}\oint_C\frac{g(\mu)}{\mu(\mu-\zeta)}{\rm d}\mu=\frac{g(\zeta)}{\zeta}-\sum\limits_{j=0}^n\frac{b_{j}}{\zeta^{j+1}}. \eqno(A.12)\]
Then, from (A.12) and (A.9), we have
\[\frac{1}{2\pi i}\oint_{\Gamma_r^-}\frac{f(k)}{k-z }{\rm d}k=f(z)-\sum\limits_{j=0}^nb_jz^j. \eqno(A.13)\]

On the other hand, in the annulus $r<|z|<R$, the Cauchy's formula implies that
\[f(z)=\frac{1}{2\pi i}\oint_{\Gamma_R}\frac{f(k)}{k-z }{\rm d}k+\frac{1}{2\pi i}\oint_{\Gamma_r^-}\frac{f(k)}{k-z }{\rm d}k. \eqno(A.14)\]
Equation (A.8) is obtained by comparing (A.13) and (A.14).

{\bf Lemma} 1 and {\bf Lemma} 2 give the proof the {\bf Theorem}.

 \end{document}